\documentclass[openacc]{rstransa}%%%%where rstrans is the template name

\begin{document}

%%%% Article title to be placed here
\title{Aether, Dark Energy, and String Compactifications}

\author{%%%% Author details
Paul K. Townsend$^{1}$}

%%%%%%%%% Insert author address here
\address{$^{1}$Department of Applied Mathematics and Theoretical Physics, 
Centre for Mathematical Sciences, \\
University of Cambridge,
Wilberforce Road, Cambridge, CB3 0WA, U.K.}

%%%% Subject entries to be placed here %%%%
\subject{Relativity, String-Theory, Cosmology}

%%%% Keyword entries to be placed here %%%%

%%%% Insert corresponding author and its email address}
\corres{Paul K. Townsend\\
\email{pkt10@cam.ac.uk}}

%%%% Abstract text to be placed here %%%%%%%%%%%%
\begin{abstract}
The 19th century Aether died with Special Relativity but was resurrected by General Relativity in the 
form of dark energy; a tensile material with tension equal to its energy density. Such a material is provided by the 
D-branes of string-theory; these can support the fields of supersymmetric particle-physics, although their energy 
density is cancelled by orientifold singularities upon compactification.  Dark energy can still arise from
supersymmetry-breaking anti-D-branes but it is probably time-dependent. Recent results on time-dependent 
compactifications to an FLRW universe with late-time accelerated expansion are reviewed. 

\end{abstract}
%%%%%%%%%%%%%%%%%%%%%%%%%%%

%%%%%%%%%% Insert the texts which can accomdate on firstpage in the tag "fmtext" %%%%%

\begin{fmtext}
\section{Introduction}

%%%% Insert A head here

When Einstein introduced his 1905  theory of Special  Relativity (SR), as it was soon to be called, he was motivated
by the need to reconcile observations implied by Maxwell's  electrodynamics for inertial observers in constant relative motion, 
in particular an invariable {\it in vacuo} speed of light $c$. The hypothesis that light waves are disturbances of a
space-filling ``aether'' was, as he put it, unnecessary \cite{Einstein}. It appears that Einstein was not motivated by the failure of Michelson 
and Morley to detect the motion of the Earth through the hypothetical aether, but SR explained their null results and
the aether hypothesis was eventually abandoned.  However, SR does not refute the aether hypothesis; it just restricts the 
aether to be some material that is Lorentz invariant, which implies that its stress-energy tensor is proportional to the
Minkowski metric. Put differently, the aether (if it exists) must be a  shear-free tensile material with a constant tension $T$ that is equal to
 its energy density ${\mathcal E}$. No such material  is known but  one may still ask whether it is theoretically possible. 

\end{fmtext}
 
 %%%%%%%%%%%%%%% End of first page %%%%%%%%%%%%%%%%%%%%%

\maketitle

Ironically, the answer to this question was implicitly provided by Einstein himself when he modified his 1915 gravitational field equations of
General Relativity (GR) to incorporate his cosmological  constant $\lambda$, with dimensions of inverse length squared \cite{Einstein2}. In modern 
notation this modified equation is 
\begin{equation}
G_{\mu\nu} +  \lambda g_{\mu\nu} = \kappa^2  \Theta_{\mu\nu}\, ,  \qquad  \kappa^2 = \frac{8\pi G_N}{c^4}\, , 
\end{equation}
where $G$ is the Einstein tensor for the spacetime metric $g$, and $\Theta$ is the matter stress-energy tensor (and $G_N$ is Newton's constant). 
However, there is an alternative interpretation of this equation, due to Lema{\^i}tre \cite{lemaitre}: we can rewrite it as the unmodified Einstein field equation
\begin{equation}
G_{\mu\nu} =  \kappa^2 T_{\mu\nu}\, , 
\end{equation}
but with the  modified matter stress-energy tensor
\begin{equation}
T_{\mu\nu} =  - \Lambda g_{\mu\nu}  + \Theta_{\mu\nu}  \qquad (\Lambda = \kappa^{-2} \lambda) \, .   
\end{equation}
This shows that the introduction by Einstein of a positive cosmological constant is equivalent to an assumption that the universe
is filled with an ideal fluid of negative pressure equal to its constant energy density: $\mathcal{E}=\Lambda$. This  ``dark energy''  
can be equivalently described as a tensile material of tension $\Lambda$ that is locally Lorentz invariant  (since $g=\eta$ in local inertial frames). 
Dark energy is therefore a kind of aether, but is it the kind that could serve as a medium 
for the propagation of electromagnetic waves? 

The short answer is ``yes but no but maybe''. The preliminary ``yes'' arises from consideration of the dynamics of a 
D3-brane of IIB superstring theory,  but this has to be integrated into a bigger picture that involves all the other fundamental forces, including gravity. 
This leads to consideration of Calabi-Yau compactifications of Type IIB superstring theory with D-branes and orientifold planes, which is a
string-theory construction of supersymmetric versions of particle physics models coupled to supergravity; a brief sketch will suffice here.
At this point the answer to our question  is ``no'' because unbroken supersymmetry doesn't allow for dark energy; it has been unavoidably 
cancelled by the orientifold planes. The introduction of anti-D-branes is one way to both break supersymmetry and introduce dark energy,
so the ``no'' becomes a ``maybe'', but here we leave the branes and proceed more phenomenologically to a review of the difficulties of incorporating  
dark energy in string-theory compactifications, and some of the ideas about how it might be achieved by allowing the compact space to be time-dependent.  

The first part of this article (the relation of branes to the aether and the connection to dark energy) was developed by the
author for  talks (unpublished) at conferences in 2005 and 2015 marking the centenaries of Einstein's works on, respectively,  SR and GR.
The last part (time-dependent compactifications to accelerating universes) is based mainly on articles from 2003-2019 written in 
collaboration with Mattias Wohlfarth, Julian Sonner  and Jorge Russo.

%%%%%%%%%%%%%%%%%%%%%%%%%%%%
%%%%%%%%%%%%%%%%%%%%%%%%
\section{Strings, Branes, Aether} 

To develop some intuition, let us imagine a one-dimensional world filled with a material of constant tension $T$, and let us suppose 
that it is embedded in a three-dimension Euclidean space. What we then have is a string of tension $T$; a 
guitar string would be an example.  According to SR, a string of mass density $\mu$ will 
have energy density ${\mathcal E} = \mu c^2$, while $T \sim 10^{-11}{\mathcal E}$ for 
a typical string on a guitar.  The tension can be increased by a tuning that stretches the string, which will also slightly decrease its mass density,
so stretching will increase the ratio $T/{\mathcal E}$ but not by much before the string breaks. So let us further imagine that this guitar-string world is 
unbreakable, and can be stretched as much as we wish; can we then make the ratio $T/{\mathcal E}$ arbitrarily large? The answer  is no because SR imposes a 
fundamental  upper limit  on this ratio. Small amplitude disturbances of the string (perhaps created by the plucking of a guitarist) will travel along it at a speed 
\begin{equation}
v= \sqrt{T/\mu} = c \sqrt{T/\mathcal{E}}\, , 
\end{equation}
but SR requires  $v\le c$, and hence
\begin{equation}\label{fund.ineq}
T\le {\mathcal E}\, . 
\end{equation}
This inequality suggests the following three-fold classification of strings:

\begin{itemize}

\item $T\ll \mathcal{E}$. These are non-relativistic strings. Our  typical guitar string  is an example, as are all strings that we could 
make with materials available to us.  \\

\item $T\lesssim \mathcal{E}$. These could be called ``relativistic'' strings. An example is the Schwarzschild black string in a 
5-dimensional (5D) spacetime. This is just the Schwarzschild black-hole solution of Einstein's 4D gravitational field equations 
`lifted' to a cyclindrically-symmetric solution of the same equations in  5D; in this case
$T= \tfrac12{\mathcal E}$ \cite{Townsend:2001rg}. \\

\item $T=\mathcal{E}$. These are ``ultra-relativistic'' strings. An example is the Nielsen-Olesen string 
solution of a Higgs-type scalar field theory \cite{Nielsen:1973cs}; the `material' that makes up the string is  a line defect 
in the scalar field analogous to the Abrikosov vortex in a superconductor.  It has a potential interpretation as a ``cosmic string'', 
and an ``effective''  (low energy) description as a Nambu-Goto string. There is also an ultra-relativistic `magnetically' 
charged black string solution  of 5D supergravity \cite{Gibbons:1994vm}; in this context the bound \eqref{fund.ineq} can be interpreted as a 
``BPS bound'' on the energy density in terms of the tension, which acquires an interpretation as the `magnetic' string-charge source for
the Maxwell-like gauge field of 5D supergravity.  Finally, there is the ``fundamental'' string of string theory, which is ultra-relativistic by hypothesis; in this case 
\begin{equation} 
T=  \frac{\hbar c}{\ell_s^2} \, , 
\end{equation}
where $\ell_s$ is the ``string length'' of string theory. 

\end{itemize}
We are not restricted to strings; the same ideas apply to branes. For waves on a $p$-brane, which has $p$ space dimensions,  SR again imposes 
the inequality \eqref{fund.ineq} but the dimensions of tension are now such that 
\begin{equation}
T = \frac{\hbar c}{\ell^{(p+1)}}
\end{equation}
for some `characteristic'  length $\ell$,  which can be interpreted as the brane's width. The $p=0$ case is degenerate because there are no waves in zero dimensions. Nevertheless,  it is natural to suppose that $T\to mc^2$ for $p=0$, where $m$ is the particle's rest-mass, and $\mathcal E \to E$, its energy; in this case  $\ell$ is the particle's Compton wavelength, and the inequality \eqref{fund.ineq} is equivalent to  $mc^2\le E$ (the standard SR formula $E=mc^2$  applies to a particle at rest).  

For an ultra-relativistic $p$-brane, the dynamics at length scales much larger than $\ell$ can depend only on the $p$-brane's tension and the geometry of the $(1+p)$-dimensional worldvolume that is swept out by its time evolution.  Let $\{\xi^\mu;\mu=0,1,\dots,p\}$ be 
local coordinates on this worldvolume; its local embedding in a $(1+n)$-dimensional Minkowski spacetime, with Minkowski coordinates $\{X^m; m=0,1, \dots,n\}$, 
is then specified by functions $X^m(\xi)$. These functions are worldvolume scalar fields for the generalization to $p$ worldspace dimensions  of Dirac's membrane action \cite{Dirac:1962iy};  in units for which $\hbar=c=1$, the Dirac $p$-brane action is
\begin{equation}
I= -T\int\! d^{p+1}\xi \sqrt{-\det g} \, , 
\end{equation}
where $g$ is the worldvolume metric induced by the Minkowski spacetime metric $\eta$:
\begin{equation}
g_{\mu\nu} (\xi) = \partial_\mu X^m \partial_\nu X^n \eta_{mn} \, . 
\end{equation}
The Nambu-Goto string action is the $p=1$ case.

As we are interested here in fluctuations of an infinite static planar $p$-brane, we shall 
write $X^m = (X^\mu, X^I)$  for $I=1,\dots, n-p$, and then choose $X^\mu$ to be the local worldvolume coordinates; i.e. 
\begin{equation}
X^\mu(\xi) = \xi^\mu\, . 
\end{equation}
This is the Monge gauge choice, and in this gauge we have 
\begin{equation}
g_{\mu\nu} = \eta_{\mu\nu} + h_{\mu\nu} \, , \qquad h_{\mu\nu} = \partial_\mu {\bf X} \cdot \partial_\nu {\bf X} \, , 
\end{equation}
where ${\bf X}$ is the Euclidean $(n-p)$-vector worldvolume field with components $X^I(\xi)$. After setting 
\begin{equation}
{\bf X} = \boldsymbol{\phi}/\sqrt{T}\, , 
\end{equation} 
we find that the Monge gauge action is 
\begin{equation}\label{Mongeact1}
I_{\rm Monge} = \int\! d^{p+1}\xi \left\{ -T - \frac12 \eta^{\mu\nu} \partial_\mu \boldsymbol{\phi} \cdot \partial_\nu \boldsymbol{\phi}  + {\cal O}(1/T)  \right\}\, , 
\end{equation}
where the  ${\cal O}(1/T)$ terms are interaction terms that are at least  quartic in derivatives of  $\boldsymbol{\phi}$.   
We now have a relativistic  field theory for scalar fields of canonical dimension propagating in a $(1+p)$-dimensional Minkowski spacetime. 
The stress-energy tensor takes the form (for ``mostly plus'' metric signature)
\begin{equation}\label{branestress}
T_{\mu\nu} = -T\eta_{\mu\nu} + \Theta_{\mu\nu}(\boldsymbol{\phi})  + {\cal O}(1/T)  \, ,  
\end{equation}
where 
\begin{equation}
 \Theta_{\mu\nu}(\boldsymbol{\phi})  = \partial_\mu \boldsymbol{\phi} \cdot \partial_\nu \boldsymbol{\phi}  - 
 \frac12 \eta_{\mu\nu} \left( \eta^{\rho\sigma}  \partial_\rho\boldsymbol{\phi} \cdot \partial_\sigma\boldsymbol{\phi} \right) \, ,  
\end{equation}
which satisfies the continuity equation $\partial_\mu \Theta^{\mu\nu}=0$ as a consequence of the linearized field equation 
$\square\boldsymbol{\phi} =0$.  The first term in \eqref{branestress} is the stress-energy tensor of the infinite static planar brane on which
the scalar fields propagate as small-amplitude fluctuations in the $n-p$ `extra' dimensions inaccessible to denizens of the brane. 

This simple example demonstrates that light-speed waves of {\it scalar} fields may have an interpretation as disturbances of an 
aether  in a way that is consistent with SR.  There is more to do before the same claim can be made for the vector fields of 
electromagnetism. An {\it ad hoc} introduction of electromagnetic fields on the brane would explain neither why 
they are confined to it nor how they might be considered fluctuations of it, but supersymmetry provides the means 
to make this ``electromagnetism on the brane'' natural  because it allows scalar fields to be connected by a symmetry to 
electromagnetic fields. This possibility is not realised by brane solitons of Minkowski space field theories,  which all have an 
effective description involving fields of spins $\le \tfrac12$   \cite{Achucarro:1987nc,Townsend:1987yy}.
A closely related fact is that soliton solutions of a maximally supersymmetric field theory preserve at most half of the 
supersymmetry, which implies that a 3-brane solution of a Minkowski space field theory can have at most N=2 4D 
supersymmetry (an N=1 example was studied in \cite{Hughes:1986fa}). So we must look to (super)gravity. 

Of most relevance here are the maximal 10D supergravity theories (IIA and IIB); these have planar static black brane solutions 
that are asymptotically flat in transverse directions \cite{Horowitz:1991cd,Duff:1991pea}. A generic black $p$-brane of this type carries a $p$-form `charge-density' of 
magnitude $T\le \mathcal{E}$, which is a source for a $(p+1)$-form potential among the supergravity fields ($p=1,3,5$  for the chiral IIB supergravity). 
The terminology `BPS' is generally used for an ultra-relativistic black $p$-brane solution with $T= \mathcal{E}$,   and a special feature of a BPS $p$-brane supergravity solution 
is that it preserves half the supersymmetry of  the 10D Minkowski vacuum. Specifically, it breaks the 10D maximal spacetime supersymmetry to a maximal 
worldvolume supersymmetry; this is possible because the $p$-form charge appears in the 10D supersymmetry algebra 
\cite{deAzcarraga:1989mza,Townsend:1997wg}. The variables relevant to the  effective low-energy dynamics of the brane are  the coefficients of `zero modes'  
of the supergravity fields in the brane background; these depend only on the brane's worldvolume coordinates and they form a multiplet of worldvolume 
supersymmetry \cite{Callan:1991ky}.  For the black 3-brane of IIB supergravity,  this is the maximally-supersymmetric (N=4) Maxwell supermultiplet \cite{Duff:1991pea}; its six scalar 
fields have the obvious interpretation as perturbations in the six transverse-space directions but these are now in the same  supermultiplet as electromagnetic fields, which are therefore as relevant to the low-energy dynamics  as the scalar fields.  But what is this dynamics? 

The ten-dimensional IIA and IIB supergravity theories are the effective theories (on length scales much  greater than $\ell_s$)  for closed strings of  the IIA and IIB superstring theories,  so one might expect $p$-branes to be a non-perturbative feature  of  Type II superstring theory, and this is required by U-duality \cite{Hull:1994ys} (as reviewed in 
\cite{Townsend:1995gp}).  Remarkably, this conclusion has  confirmation from string perturbation theory: Polchinski showed that the Type II superstring theories include not only closed strings but also open strings with endpoints on fixed $p$-planes \cite{Polchinski:1995mt}, called ``D-branes''  (D$p$-branes if we want to specify $p$) because of the Dirichlet boundary conditions at the string endpoints \cite{Dai:1989ua}. They are branes because the massless modes of open strings attached to them can be interpreted as  fluctuations of the fixed $p$-plane; the D$p$-brane tension is  \cite{Polchinski:1995mt}
\begin{equation}
T_{\rm Dp}= \frac{\hbar c}{g_s\ell_s^{p+1}}\, , 
\end{equation}
where $g_s$ is the dimensionless string coupling constant ($g_s\ll1$ in string perturbation theory).  Notice that the tension 
becomes infinite in the $g_s\to0$ limit. This is a reflection of the fact that the undisturbed 
D$p$-brane fills a {\it fixed} $p$-plane in string perturbation theory, but that is precisely the circumstance of interest here since it 
provides us with a Minkowski `vacuum' on which the open-string massless modes propagate. Moreover, the effective action for these 
fields (on length scales much larger that $\ell_s$) can  be found  from superstring perturbation theory; the result is a generalisation of a 1985 result of Fradkin and Tseytlin \cite{Fradkin:1985qd} who showed that the effective action for slowly  varying massless fields  of the open bosonic string  is (a higher-dimensional version of) the Born-Infeld action for non-linear electrodynamics \cite{Born:1934gh}. The generalisation  fuses the Born-Infeld action and the  Dirac $p$-brane action into the  Dirac-Born-Infeld (DBI) action \cite{Leigh:1989jq}. 

For the IIB D3-brane the massless open-string modes form an N=4 Maxwell supermultiplet, in agreement with  the supergravity results. The DBI action for the bosonic fields of this supermultiplet  (i.e. omitting the fermionic fields required by supersymmetry) is (again for $\hbar=c=1$)
\begin{equation}
I = - T_{\rm D3} \int\! d^4\xi \sqrt{-\det\left(g+ \ell_s^2 F\right)} \, , 
\end{equation}
where $F$ is the antisymmetric worldvolume electromagnetic field-strength tensor.  In the Monge gauge for which 
\begin{equation}
g_{\mu\nu} = \eta_{\mu\nu} +  \ell_s^2\partial_\mu \boldsymbol{\phi} \cdot \partial_\nu \boldsymbol{\phi}\, , \qquad 
\boldsymbol{\phi} = \{ \phi^1, \dots,\phi^6\}\, , 
\end{equation}
the DBI action takes the form 
\begin{equation}
I_{\rm Monge} = g_s^{-1} \int\! d^4\xi \left\{ -\ell_s^{-4} - \frac12 \eta^{\mu\nu} \partial_\mu \boldsymbol{\phi} \cdot \partial_\nu \boldsymbol{\phi}  
- \frac14 \eta^{\mu\rho}\eta^{\nu\sigma} F_{\mu\nu} F_{\rho\sigma} + {\cal O}(\ell_s^2)  \right\} \, , 
\end{equation}
which may be compared with \eqref{Mongeact1}.  Apart from the overall factor of $1/g_s$ and the omitted fermion fields, the difference is that we now have 
Maxwell's electrodynamics in addition to scalars. Electric 
charges (not included in the above action) have a higher-dimensional interpretation as the endpoints  of fundamental IIB strings, and magnetic charges have a similar interpretation as end-points of IIB D-strings (i.e. D1-branes). 

Leaving aside many issues to focus on the fact that we now have an interpretation of electromagnetic waves as disturbances of an aether consistent with SR,  let us consider whether the tension of this D3-brane aether can be identified with the cosmological constant; i.e. can we have $T_{\rm D3}= \Lambda$?  The answer is no because in BI electrodynamics there is a maximum value of the electric field (this was its motivating feature) and the BI parameter $T$ is the corresponding maximum energy density, but atomic physics experiments put a lower bound on any hypothetical maximum  value of  the electric field \cite{Soff}, and this is equivalent to the bound $T\gtrsim10^{33} J/m^3$. In contrast, cosmological observations yield $\Lambda \sim 10^{-10} J/m^3$, a discrepancy of more than 40 orders of magnitude! This is not a problem for BI theory  because we can subtract a constant from the vacuum energy density (and this is usually done) but this {\it ad hoc} subtraction is contrary to the spirit of  the DBI action. Let us put this problem aside for the moment to consider another one.

%%%%%%%%%%%%%%%%%%%%%%%%%%%%%
\subsection{Gravitational aether?} 

We need a theory not just of electromagnetism but of all the fundamental interactions. There is no difficulty in getting 
other spin-1 gauge theories ``on the brane''. For example, the effective low-energy description of a parallel stack of $N$ coincident 
D3-branes is a $U(N)$ gauge theory \cite{Witten:1995im}, but this leaves out gravity. One may wonder whether there is some other 
kind of brane on which a massless spin-2 field can be trapped.  There are  reasons to think that this is not possible; a review 
of the difficulties, and an interesting potential way to overcome them, may be found in \cite{Crampton:2014hia}.  

{\it A priori}, it is hard to see how the dynamics implied by a presumed Einstein-Hilbert term ``on the brane'' could be compatible with a brane interpretation. 
One difficulty is that there is no coordinate-independent {\it local} definition of energy in GR, which is why energy loss due to gravitational radiation can only be understood in the context of appropriate asymptotic boundary conditions, as first appreciated by Trautman \cite{Trautman:1958zdi}; coordinate transformations can move the energy around like a bump under the carpet that can be flattened locally but not globally. This would appear to rule out any interpretation of gravitational waves as fluctuations of a `gravitational' aether. 

This conclusion is supported by the 't Hooft-Susskind  principle of holography that was suggested by black-hole physics \cite{tHooft:1993dmi,Susskind:1994vu}. This states that the number of quantum gravitational degrees of freedom in a given region grows as the area of the region's boundary, but the number of degrees of freedom associated with quantised fluctuations  of a gravitational aether would grow with the volume (by hypothesis, essentially)  and these two growth rates are generally very different; an exception is 
anti-de Sitter spacetime because the volume of  a ball in  hyperbolic space grows asymptotically (and in suitable units) at the same rate as the area of its boundary (a fact that is essential for the applicability of the holographic principle in anti-de Sitter spacetime \cite{Susskind:1998dq}).

Nothing said above implies the impossibility of a gravitational aether. An ``Einstein-Aether'' modification of GR  has been proposed \cite{Eling:2004dk}, for  phenomenological reasons reviewed in \cite{Jacobson:2007veq}, but it does not appear to arise from any larger framework such as string/M-theory that could provide a consistent  quantum theory. In other words, it may be in the ``swampland'', which is the currently popular term for phenomenological theories that have no ``UV completion''; the topic was recently reviewed in  \cite{Palti:2019pca}.

%%%%%%%%%%%%%%%%%%%%%%%%%%%%%
%%%%%%%%%%%%%%%%%%%%%%%%%%%%
\section{String-theory and dark energy}

The standard, Kaluza-Klein, way to get a theory with gravity in four-dimensions from one that is initially formulated 
in a higher dimension, such as superstring theory, is to look for a solution for which the `extra' dimensions are compact. For
IIB superstring theory we must compactify six dimensions. In the remaining three space dimensions, and at distinct points in the compact space, 
we may place stacks of D3-branes. More generally, we may have stacks of D(3+2k)-branes wrapped on $2k$-cycles of the compact 
manifold for $k=0,1,2$.  In practice, the compact space is chosen to be a Calabi-Yau manifold, which yields an effective N=2 4D supergravity, and the $2k$-cycles for $k>0$ are chosen to be calibrated surfaces, breaking N=2 supersymmetry to N=1.  In this way we might hope  to have a physical 3-space filled with stacks of D-branes, some wrapped over cycles of the compact manifold, with each stack separated by some distance in the compact manifold. However,  the total D$p$-brane charge (for each $p$) must be zero unless there are singularities in the compact space that can act as sinks for the (p+2)-form electromagnetic-type fields sourced by the D$p$-branes. 

In string theory there are allowed singularities of this type, orientifolds, which cancel both the D-brane charge {\it and} the energy density,  allowing a compactification to 4D Minkowski space. In this way, one can `engineer' quasi-realistic  N=1 supersymmetric extensions of the standard model of particle physics,  as reviewed in \cite{Blumenhagen:2006ci}, with
the particles arising as massless modes of open IIB superstrings with ends on D-branes. In the 4D effective theory, valid on length scales much greater than $\ell_s$, the fields for these particles are coupled to N=1 supergravity, but the graviton and gravitino are massless modes of closed IIB superstrings that propagate freely in the full 10D spacetime as particles
of (quantum) size $\ell_s$. They interact weakly with the particles on the branes because $\ell_s$ is much  greater (for $g_s\ll 1$) than the width of any of the D-branes. 

The stability of these constructions depends partly on the assumption of preservation of N=1 4D supersymmetry, which allows the inter-brane forces to cancel, and on fluxes in the compact space that allow a stabilisation of moduli (parameters of the compact space, such as its scale).  We still have an interpretation of particles as quanta associated to fluctuations of branes, which collectively serve as an aether,  but there is no longer any connection to the cosmological constant. This resolves the discrepancy with atomic physics experiments noted earlier, but  it also leaves us without dark energy; to incorporate it we need some supersymmetry-breaking modification, which is needed anyway to arrive at the non-supersymmetric  standard model of particle physics.  

One possibility that maintains a connection with branes is to add anti-branes; we should then expect instabilities although an influential 2003 model incorporating anti-D3-branes was argued to have a (metastable) de Sitter vacuum \cite{Kachru:2003aw,Polchinski:2015bea}; this is the late time universe to which our Universe must approach if the cosmological constant is really constant, as it is within observational bounds \cite{Planck:2018vyg}. The de Sitter universe may be viewed as a flat FLRW spacetime with an exponential scale factor:
\begin{equation}
ds^2 = -dt^2 + S^2 d{\bf x}\cdot d{\bf x} \, , \qquad S= e^{Ht} \, , \quad H= \sqrt{3c^2\lambda}\, . 
\end{equation}
This universe is expanding at a rate determined by the  ``Hubble constant''  $H=\dot S/S$, which really is constant in this case.  The expansion is accelerating at a rate determined by  $\ddot S/S$, which is the constant $H^2$ in this case.  The accelerated expansion dilutes all but the dark energy density and smoothes out any inhomogeneities, so the de Sitter universe is a late-time ``attractor'' solution of the Einstein field equations. 

However, there is mounting evidence, e.g. \cite{Sethi:2017phn,Danielsson:2018ztv,Obied:2018sgi},  that  all ``de Sitter compactifications'' are part of the swampland. This can be seen as a conjectural generalisation of an earlier no-go theorem that rules out non-singular de Sitter compactification solutions of the (higher-dimensional) Einstein field equations if the compact space (and warp factor in the case of ``warped'' compactifications) are time-independent {\it and} the stress-energy tensor satisfies the strong-energy condition (SEC)  \cite{Gibbons:1984kp}.  In the context of the effective low-energy 10D or 11D supergravity theories of string/M-theory this result was rediscovered in \cite{Maldacena:2000mw} , where it was also extended to cover one exceptional case for which the SEC does not hold (massive IIA). It should be appreciated that the SEC is not required by basic  physical principles; its significance comes from the fact that it is satisfied by the effective low-energy 10D or 11D supergravity theories of string/M-theory. More recently it was shown that  {\it time-dependent} compactifications to the de Sitter universe on  non-singular compact spaces can also be ruled out  if the higher-dimensional stress-energy tensor satisfies both the SEC and the null energy condition (NEC) \cite{Russo:2019fnk} (although the dominant energy condition (DEC)  was stated as a premise, only the weaker NEC was actually used in the proof).  The tension between string/M-theory and the astronomical observations that indicate a constant dark-energy density is therefore far from having been resolved, despite two decades of  effort. 

From a purely 4D  perspective, the effect of `extra' inaccessible dimensions is to provide a variety of matter fields and a potential $V$ for any 
scalar fields associated to moduli (of the compact space or D-brane configurations) and relative positions of anti-branes (see e.g. \cite{Polchinski:2015bea,Sethi:2017phn}). 
The Friedmann equations found from those of GR by requiring spacetime homogeneity and isotropy then yield the possible FLRW universes obtainable
by the compactification considered.  In this context, the absence of any de Sitter compactification is equivalent to the statement that $V$ has no stationary points 
at which $V>0$, which means that any dark-energy density must be time-dependent. This might be compatible with observations if the time-dependence is sufficiently slow, and this possibility motivates an exploration of compactifications to FLRW universes other than de Sitter for which there is an accelerated cosmic expansion, i.e. $\dot S>0$ and $\ddot S>0$. 
It is not difficult to find simple examples of string/M-theory compactifications on {\it time-dependent} compact spaces that lead to flat FLRW cosmologies with a short period of accelerated expansion, i.e. ``transient acceleration'' (see e.g. \cite{Townsend:2003fx,Ohta:2003ie,Wohlfarth:2003kw}) but is {\it late-time} acceleration possible?  

The answer to this question is determined by properties of the 4D scalar potential $V$.
An instructive, and much studied, example is  a single scalar field $\sigma$ with the energy density
\begin{equation}
\mathcal{E} = \frac12 \dot\sigma^2 + V(\sigma) \, , \qquad V(\sigma) = \Lambda e^{- 2\alpha \sigma} \, , 
\end{equation}
for positive constant $\alpha$. We may restrict attention to flat universes since this is implied by late-time accelerated expansion, and for this simple case 
all flat FLRW universes can be found exactly \cite{Russo:2004ym} (for not too large $\alpha$).
Typical solutions correspond to motion up and then down the potential, and accelerated cosmic expansion occurs around the stationary 
point of this motion, when $V$ is approximately constant \cite{Emparan:2003gg}. There is then a transition to a late-time ``scaling'' solution with 
\begin{equation}
S\sim t^{\eta(\alpha)} \, , \qquad \sigma \sim {\alpha}^{-1}  \ln t \, . 
\end{equation}
The late-time expansion is accelerating only if $\eta\ge1$, which requires $\alpha\le1$. 
Although an exponential potential is special, a late time attractor solution yielding an accelerating FLRW universe for generic positive multi-scalar potential $V(\boldsymbol{\sigma})$  will be such that \cite{Townsend:2004zp} 
\begin{equation}
S\sim t^{\eta(\boldsymbol{\alpha})}\, , \qquad \lim_{t\to \infty} \left[\frac{\partial (\ln V)}{\partial \boldsymbol{\sigma}(t)}\right] 
= -2 \boldsymbol{\alpha} \, , \qquad |\boldsymbol{\alpha}| \le 1\, . 
\end{equation} 

There is no generally accepted string/M-theory compactification to an FLRW universe with late-time accelerating expansion, but it was recently shown that this
is {\it not} ruled out by the premises of the no-go theorems summarised above; in particular, it is permitted by the higher-dimensional SEC and DEC 
combined \cite{Russo:2018akp}.  This result was found by focusing on cosmological compactifications to FLRW spacetimes that could serve as late-time power-law attractor solutions, and then using the higher-dimensional Einstein field equations to determine the stress-energy tensor in terms of $\eta$ and another exponent $\xi$ 
controlling the change of scale of the compact space. The SEC limits the values of $(\eta,\xi)$ to the interior of an ellipse, and the acceleration condition $\eta\ge1$
further limits them to a chord segment of the ellipse; the DEC (and hence the NEC) is then automatically satisfied. 

In all these cases,  $\mathcal{E} \sim t^{-2}$ at late times, so the later the time the lower the dark energy density, and if $t_0$ is a late time then the further change in 
$\mathcal{E}$ over a period $\Delta t << t_0$ will be small.  However, the 4D stress-energy tensor at late times was found in \cite{Russo:2018akp} to be an ideal fluid with a pressure equal to $w\mathcal{E}$, where the constant $w$ satisfies 
\begin{equation} \label{SECbound}
- \tfrac12 < w < -\tfrac13\, . 
\end{equation}
The upper bound is needed for accelerating  expansion of the 4D FLRW universe. The lower bound comes from the SEC that the
$D$-dimensional stress-energy tensor is assumed to satisfy.  By contrast, the constant dark energy implicit in a late-time de Sitter universe 
is an ideal fluid with $w=-1$ (negative pressure equal to the energy density).  Observations are consistent with $w=-1$ but {\it not} with  $w>-\tfrac12$, so
we need to re-examine the assumptions leading to \eqref{SECbound}.

String-theory introduces higher-derivative corrections to the Einstein field equations. These could be accommodated in the analysis of \cite{Russo:2018akp} by rewriting 
the corrected field equations as the uncorrected equations with a corrected stress-energy tensor, which might  violate the SEC. However,  if this {\it is} the effect of string-theory corrections  then one of the major obstacles to finding a de Sitter compactification, and hence $w=-1$, is removed.  But it is precisely the difficulties that have been encountered in trying to achieve this goal that have motivated the conjecture that de Sitter compactifications belong to the swampland. For this reason, it seems unlikely that the higher-derivative corrections of string theory can help. There remains the possibility that singular compactifications allowed by string-theory (but not supergravity) will allow $w<-\tfrac12$, but it is unclear how this could be achieved.  The idea that dark energy can emerge from string-theory as a byproduct of cosmological compactifications to FLRW universes with an accelerated cosmic expansion powered by  some ideal fluid with $w\lesssim 1$ has now run into difficulties, which go beyond those mentioned here; see e.g. \cite{Krishnan:2021dyb}. 

Despite these difficulties, there is one result that may turn out to be significant.  A problem with time-dependent compactifications is that an expanding 4D FLRW universe typically requires an expanding compact space, which could lead to  unacceptable time-dependence of parameters of the effective four-dimensional theory. In addition, eternal expansion would imply an ultimate decompactification: the Kaluza-Klein (non-zero) modes will become directly observable at some ``decompactification time''  $t_{\rm decomp}$.   A remarkable, and surprising,  corollary of the analysis of \cite{Russo:2018akp}  is that accelerating expansion of the FLRW universe implies {\it decelerating} expansion  of the compact space. This allows $t_{\rm decomp}$ to be easily much greater than the age of our Universe, and may reduce time-dependence of four-dimensional 
parameters to an acceptable level. 

%%%%%%%%%%%%%%%%%%%%%%%%%
\subsection{Recurrent acceleration}

So far we have focused on the {\it late-time} fate of the Universe, mainly because a constant dark energy density implies that this will be a de Sitter universe. This
is likely to be unrealisable by any cosmological compactification of string/M-theory, so it is natural to seek alternative possibilities that could be 
both realisable and compatible with observations. As noted above in the context of an effective 4D theory with a positive potential $V(\boldsymbol{\sigma})$ 
without stationary points, $\mathcal{E}(t) \le V_{\rm max}$ for any given flat FLRW solution, where $V_{\rm max}$ is a maximum value of $V$ that is reached when 
$\dot{\boldsymbol{\sigma}}={\bf 0}$. Around this time, $\mathcal{E}(t)$ is approximately constant, and the scale factor $S(t)$ grows exponentially, as in the de Sitter universe. 
However, this de Sitter-like phase is transient.

For the simple one-scalar model with exponential potential discussed earlier, there is only one phase of transient acceleration if $\alpha\equiv \tfrac12 |V'|/V >1$.
This is likely to remain true for arbitrary one-scalar positive potentials unless $\tfrac12|V'|/V<1$ for some values of $\sigma$  because this is needed for 
late-time acceleration. In any case, there is no known {\it one-scalar} example found by cosmological compactification (and this fact is likely related to the 
de Sitter swampland conjecture that requires $|V'|/V>C$ for some (unknown) constant of order $1$ \cite{Obied:2018sgi}).  
However, for a simple dilaton-axion  model with 
\begin{equation}
\mathcal{E}= \frac12 \left(\dot\sigma^2 + e^{-2\beta\sigma}\dot\chi^2\right) + \Lambda e^{-2\alpha\sigma}\, , 
\end{equation}
there can be multiple phases of accelerating expansion when $\alpha>1$ (and even $\alpha\gg1$) if the dilaton-axion coupling 
constant $\beta$ is sufficiently large; this is the phenomenon of ``recurrent acceleration'' \cite{Sonner:2006yn}. A large dilaton-axion coupling will
rapidly convert dilaton kinetic energy into axion kinetic energy, which slows the descent down the exponential potential;  the energy density 
is now decreasing more slowly, sufficiently to drive an accelerated expansion. But then the axion kinetic energy is reconverted into dilaton kinetic energy, 
which speeds up the descent down the potential; the energy density is now falling rapidly and the expansion is decelerating.  After a few cycles the late-time (accelerating or decelerating) scaling solution  is approached and $\mathcal{E}\sim t^{-2}$.  The first phase of acceleration is de-Sitter-like because, momentarily  $\dot\sigma =\dot\chi=0$, 
whereas subsequent phases of acceleration are presumably closer to power-law accelerated expansion. 

More generally, in a universe with recurrent acceleration, the duration of a first, de-Sitter-like, phase of accelerated expansion will be prolonged by the conversion of potential energy into kinetic energy of fields on which $V$ has no (or lesser) dependence.  As pointed out in \cite{Russo:2018akp}, there is a simple mechanical analogy to a disc rolling down a hill; potential energy is  converted into a combination of translational kinetic energy (which determines the rate of descent) and rotational  kinetic energy (which has no effect on the rate of descent), so the larger the  moment of inertia of the disc the more its descent will be slowed. Large moment of inertia in this mechanical model corresponds to large coupling constant $\beta$ in the dilaton-axion model.  Potentially, our Universe could be in a phase of ``suspended de-Sitter-like'' expansion, which will later turn to decelerating expansion; in this case, the late-time future will remain open, and probably unknowable.

%%%%%%%%%%%%%%%%%%%%%%%%%%%%%%%%%
%%%%%%%%%%%%%%%%%%%%%%%%%%%%%%%
\section{Summary and outlook} 

Dark energy, which is about 70\%  of the total energy of the Universe,  is possibly the simplest form of energy. Astronomical observations suggest that its density is 
constant, in which case it can be identified as Einstein's cosmological constant, implying an evolution towards a de Sitter universe. 
An equivalent interpretation  is as a space-filling shear-free tensile material with tension equal to its 
energy density, i.e. $T=\mathcal{E}$.  It has been pointed out here that such a material  is a candidate for an SR-compatible aether, and that 
branes provide a realisation of  this idea since their fluctuations propagate as light-speed waves, but only ``on the brane''. 
In the case of D-branes of superstring theory,  these include electromagnetic waves. 

This idea extends to all the non-gravitational forces in the context of  constructions of supersymmetric  particle-physics models by compactifications 
of IIB superstring theory with D-branes on Calabi-Yau manifolds, since the particles are then  quantised fluctuations of (stacks of) D-branes, which 
are well-separated in the compact space and which interact weakly with supergravity fields. However, the  D-brane energy density of the 4D effective theory 
has now been cancelled by orientifold singularities of the compact space, and there is therefore no dark energy.  

The addition of anti-branes can raise the energy density,  breaking supersymmetry and potentially producing a constant dark-energy density, 
implying a late-time de Sitter universe. However, it now seems likely that string-theory scenarios for compactification to a de Sitter universe belong 
to the `swampland'; this is a conjectured extension of earlier no-go theorems based on an improved understanding of how many attempts to evade 
them have failed. If correct,  string-theory predicts a time-dependent  dark energy density, and hence some alternative late-time universe.  

All FLRW universes that undergo accelerated expansion can be late-time attractor solutions of the Einstein field equations. The constant energy density 
that drives the de Sitter expansion is an ideal fluid with pressure to energy-density ratio $w=-1$, which is equivalent to $T=\mathcal{E}$. For $w>-1$, equivalently
$T<\mathcal{E}$, we get other `scaling' solutions with a power-law expansion that is accelerating if  $w<-\tfrac13$.  There is no known string-theory cosmological 
compactification to {\it any} FLRW universe with late-time accelerated expansion, but one may still ask whether the premises of the de Sitter no-go theorems exclude it. 
They do not exclude it, but they do impose the bound $w>-\tfrac12$, which is not compatible with observations. 

One remaining possibility exploits the fact that {\it transient} de Sitter-like acceleration is generic for any compactification  that leads to an effective
4D theory with a positive potential $V$. This is easily achieved; the no-go theorems require only that $V$ has no stationary point for $V>0$,  and the de Sitter 
swampland conjecture imposes bounds on the magnitude of the gradient of $V$. Cosmological evolution  involves rolling up the potential to a 
maximum value, and then a rolling down; this implies a moment at which the total energy density $\mathcal{E}$ is stationary and this drives a transient 
phase of de Sitter-like expansion. Although this  may be too short, generically, its duration can be extended in multi-scalar models that exhibit recurrent 
acceleration because the mechanism underlying this phenomenon is one that slows the fall of $\mathcal{E}$ from its maximum value. 
It remains to be seen whether this slowing is sufficient. 

Whatever the resolution of tension between string theory and the observed accelerated expansion of our Universe, one lesson to be learned 
is  surely that it would be naive to suppose of any cosmological compactification of string theory that the compact space  is time-independent. 
 However, its time dependence could easily lead to an unacceptably large time-dependence of parameters of the effective 
four-dimensional  theory and a possible decompactification.  One feature of accelerated expansion is that it can freeze the time evolution of the compact space; in the analysis leading to the lower bound on $w$ mentioned above, this was due the fact that an accelerating expansion of the 4D universe required a decelerating expansion of the compact space. 

Dark energy may be simple but it is not so simply accommodated by string/M-theory, which in most other respects is still our 
best collection of ideas for a unified theory of particle physics with quantum gravity. It may be time to look for new ideas!

\funding{This work was partially supported by STFC consolidated grant ST/P000681/1}

\ack{I am grateful to Eoin Colg\'ain and Jorge Russo for helpful communications and conversations.} 

 %%%%%%%%%% Insert bibliography here %%%%%%%%%%%%%%

\end{document}